\documentclass[a4paper, twocolumn, superscriptaddress,prl]{revtex4}  

\usepackage{graphicx}
\usepackage{amssymb,amsfonts,amsmath}
\usepackage{nicefrac}
\usepackage[rightcaption]{sidecap}
\usepackage{color}
\usepackage{ulem} 

\begin{document}

\title{Atomic Quadrupole Moment Measurement Using Dynamic Decoupling}

\author{R. Shaniv}

\author{N. Akerman}

\author{R. Ozeri}
\affiliation{Department of Physics of Complex Systems, Weizmann Institute of Science, Rehovot, Israel}

\begin{abstract}
We present a method that uses dynamic decoupling of a multi-level quantum probe to distinguish small frequency shifts that depend on $m^2_{j}$, where $m^2_{j}$ is the angular momentum of level $\left|j\right\rangle$ along the quantization axis, from large noisy shifts that are linear in $m_{j}$, such as those due to magnetic field noise. Using this method we measured the electric quadrupole moment of the $4D_{\frac{5}{2}}$ level in $^{88}Sr^{+}$ to be $2.973^{+0.026}_{-0.033}\, ea_{0}^{2}$. Our measurement improves the uncertainty of this value by an order of magnitude and thus helps mitigate an important systematic uncertainty in $^{88}Sr^{+}$ based optical atomic clocks and verifies complicated many-body quantum calculations.
\end{abstract}

\maketitle

Increasing the coherence time of quantum superpositions is an ongoing research topic. From the quantum information point of view, prolonging the coherence time allows for the performance of a larger number of coherent operations, making more complex quantum computation algorithms possible \cite{divincenzo1995quantum} as well as longer-living quantum memory \cite{langer2005long}. From a metrology perspective, decoherence poses a time limit on the coherent  evolution of a quantum probe when used to measure some physical quantity, thereby limiting the accuracy with which this quantity is evaluated. \par

Several methods were developed and demonstrated to increase the coherence time of quantum systems. Examples include decoherence free subspaces (DFS) \cite{DFS-review,kielpinski2001decoherence}, in which states are invariant under the effect of the environment, quantum error correction codes (QECC) \cite{shor1995scheme,steane1996error}, in which an error can be detected by a measurement that does not reveal information on the encoded superposition and can be subsequently corrected, and dynamic decoupling (DD) methods in which a quantum superposition is spectrally separated from its  environment (a review of DD methods can be found in \cite{gordon2007universal}). The optimal spectrum of qubit modulation depends on the details of noise. While DD methods are not as general as QECC and cannot mitigate the effect of any type of noise, their implementation is much simpler as it does not require the entanglement of multi-qubit arrays and relies solely on single qubit operations. All the methods above have been suggested or used to improve quantum metrology as well \cite{hall2010ultrasensitive,Single-Spin_G_de_Lange,kotler2011single,ozeri2013heisenberg,kotler2013nonlinear,baumgart2014ultrasensitive,kotler2014measurement,kessler2014quantum}. In the context of metrology a method is chosen such that on top of noise suppression the measured signal is coherently accumulated. \par

In this work we present a new DD method which decouples a superposition of atomic levels in a single trapped $^{88}Sr^{+}$ ion from external magnetic field noise while measuring the electric quadrupole shift induced by an electric field gradient. In contrast with usual two-level quantum probes, here we take advantage of the six-fold $4D_{\frac{5}{2}}$ Zeeman manifold of equidistant levels. Using this method we measured the quadrupole moment of the $4D_{\frac{5}{2}}$ level in $^{88}Sr^{+}$ to be $2.973^{+0.026}_{-0.033}\, ea_{0}^{2}$, where $e$ is the electron charge and $a_{0}$ is Bohr radius. \par

The quadrupole moment of an atomic state is a measure of its deviation from perfect spherical symmetry. While ground state $S$ orbitals are spherically symmetric, higher levels, such as states in the $D$ manifold have a finite electric quadrupole moment. This electric quadrupole moment couples to an electric field gradient across the atomic wavefunction. The resulting energy shift is usually small in comparison to other shifts, e.g. Zeeman shift, due to the small atomic length scale over which the electric field has to change. However small, this is nevertheless an important systematic shift in ion-trap optical atomic clocks \cite{poli2014optical}. Here, electric field gradients are inherent to the trap and induce typical shifts to optical electric-quadrupole clock transitions on the order of 10's-100's Hz which constitute fractional frequency shifts on the order of $10^{-13}$. While the electric field gradient at the ion position can be estimated by measuring the trap spring constant, evaluating the resulting shift requires a reliable knowledge of the quadrupole moment of the levels involved. \par

We performed our measurement on a single $^{88}Sr^{+}$ ion trapped in a linear Paul trap. The $S_{1/2}\rightarrow D_{5/2}$ electric quadrupole optical transition of this ion is investigated as a possible atomic time standard in several experiments \cite{margolis2004hertz,dube2013evaluation}.
 The quadrupole moment of the $4D_{\frac{5}{2}}$ level in $^{88}Sr^{+}$ was previously measured \cite{barwood2004measurement}. Measurements of the D level quadrupole moment of $^{199}Hg^{+}$ \cite{oskay2005measurement} and $^{171}Yb^{+}$ \cite{schneider2005sub} were also published. In addition, a method to measure the quadrupole moment of $^{40}Ca^{+}$ using an entangled superposition of two ions was experimentally demonstrated \cite{roos2006precision}. Our new method yields a measured value for the quadrupole moment in the $4D_{\frac{5}{2}}$ level in an $^{88}Sr^{+}$ ion with about ten times smaller uncertainty in comparison to the previous measurement, and requires a single ion only. \par

The relevant levels for our experiment are the two-fold $5S_{\frac{1}{2}}$ ground level $|S, m_j\rangle$, and the six-fold $4D_{\frac{5}{2}}$ level  $|D, m_j\rangle$ shown schematically on Fig. 1. A constant magnetic field of typically $B=3\times10^{-4}\,\mbox{T}$, splits the different Zeeman states in both levels. This magnetic field is generated by three Helmholtz coils in three orthogonal axes. The relative current between the coils determines the magnetic field direction with respect to the trap quadrupole axis.
The Zeeman frequency splitting between two adjacent levels in the $4D_{\frac{5}{2}}$ manifold  $\Delta f_{Z}$ is typically around $5\,\mbox{MHz}$. The ion $5S_{\frac{1}{2}}\rightarrow4D_{\frac{5}{2}}$ transition is addressed with a 674 nm narrow linewidth laser, with spectral fast line-width of 100 Hz over several minutes \cite{akerman2015universal}. An RF electrode located two mm away from the ion trap center generates oscillating magnetic fields at the Zeeman splitting frequency leading to spin rotations. The orientation of the trap electrodes, the laser $\vec{k}$ vector and the quantization magnetic field are shown in Fig. 1a. Ion state detection is performed using state selective fluorescence on a strong dipole allowed transition at 422 nm \cite{keselman2011high}. More details about the experimental apparatus can be found in \cite{akerman2012quantum}. \par
 While the $5S_{\frac{1}{2}}$ electronic wavefunction is spherically symmetric, and therefore has no electric quadrupole moment, states in the $4D_{\frac{5}{2}}$ level have a quadrupole charge distribution, and therefore an electric field gradient across these wavefunctions will shift their energy. Since the ion is held in a linear Paul trap, it is held in a constant (DC) electric field gradient induced by the trap electrodes. The electric DC potential generated by these electrodes is well approximated by,
\begin{equation}
U=\frac{1}{4}\frac{dE_{z}}{dz}(x^{2}+y^{2}-2z^{2}),
 \end{equation}
 where $z$ is the direction of the trap axis, and the resulting shift of state $\left|D,m_{j}\right\rangle$ level is given by \cite{itano2000external},
 \begin{equation}
\Delta\nu_{Q}=\frac{1}{4h}\frac{dE_{z}}{dz}\Theta\left(D,\frac{5}{2}\right)\frac{35-12m_{j}^{2}}{40}\left(3\mbox{cos}^{2}\left(\beta\right)-1\right).
\end{equation}
Here, $\frac{dE_{z}}{dz}$ is the electric field gradient along the quadrupole axis, $\Theta\left(D,\frac{5}{2}\right)$ is the $4D_{\frac{5}{2}}$ level quadrupole moment and $\beta$ is the angle between the quantization axis set by the magnetic field and the quadrupole axis set by the trap geometry. In our trap this shift is on the order of 50 $\mbox{Hz}$ resulting from typical $\frac{dE_{z}}{dz}\sim100\mbox{\ensuremath{\frac{V}{mm^{2}}}}$ . This small shift is usually overshadowed by the much larger time variation in Zeeman shifts, typically on the order of a few kHz in several $\mbox{msec}$. In order to measure the quadrupole shift, we developed a DD sequence that employs the equidistant $4D_{\frac{5}{2}}$ Zeeman sub-levels to decouple our ion from effects linear with $m_{J}$, while leaving it susceptible to shifts that depend on $m_{J}^{2}$. Unlike most previous measurements of clock transition quadrupole moments, our measurement protocol involves only the $D_\frac{5}{2}$ levels and radio frequency (RF) transitions and no optical transitions. Therefore, many systematic effects that affected previous quadrupole shift measurements \cite{barwood2004measurement,oskay2005measurement,schneider2005sub} such as ac Stark shifts, second-order Doppler effects and their dependence on micromotion, as well as limitations imposed by the clock transition laser phase noise, are irrelevant. \par

\begin{figure}
\centering
\includegraphics[width=5.6cm]{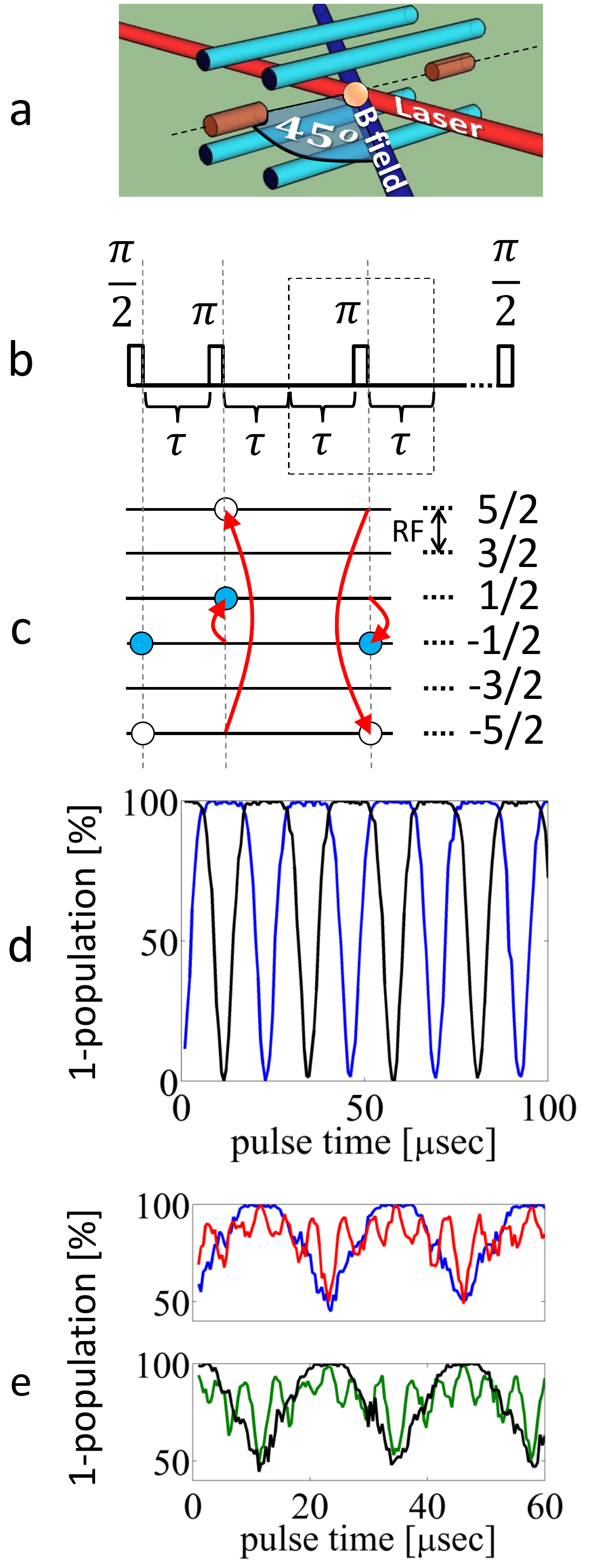}
\caption{Layout of the experiment and pulse sequence scheme.\\
\textbf{(a)} Scheme of the trap structure and orientation with respect to the laser $\vec{k}$ vector and the magnetic field directions. The trap axis is shown by the dashed line passing through both endcaps. The laser direction is at a right angle with respect to the trap axis, the magnetic field direction can be scanned and is shown here to be at $45^{o}$ angle with respect to both the laser and the trap axis. \textbf{(b)} Pulse sequence for the measurement of the quadrupole shift. The dashed frame marks the sequence building-block which is being repeated. \textbf{(c)} The definition of an RF $\pi$ pulse. The blue and white circles mark the two amplitudes, and the numbers on the right column denote the different $m_{J}$ values. The black two-sided arrow marks the RF frequency used to drive the transition. \textbf{(d),(e)} Rabi oscillations on the six-fold $4D_{\frac{5}{2}}$ energy level for initial states $\left|\psi_{i}\right\rangle =\left|D,-\frac{5}{2}\right\rangle$ and $\left|\psi_{i}\right\rangle =\frac{1}{\sqrt{2}}\left(\left|D,-\frac{5}{2}\right\rangle +\left|D,-\frac{1}{2}\right\rangle \right)$ respectively. The vertical axis shows the population outside the detected sub-level. Blue, red, green and black lines correspond to the population in the levels $m_{J}=-\frac{5}{2}$, $m_{J}=-\frac{1}{2}$, $m_{J}=\frac{1}{2}$ and $m_{J}=\frac{5}{2}$ respectively. Both upper and lower plots in (e) correspond to the same experiment, separated for clarity.}
\end{figure}

Our experimental sequence is shown in Fig. 1b. We begin the sequence by optically pumping the ion to the $\left|S,-\frac{1}{2}\right\rangle$ state. We then apply a $\frac{\pi}{2}$ pulse on the $\left|S,-\frac{1}{2}\right\rangle \rightarrow\left|D,-\frac{5}{2}\right\rangle$ transition followed by a $\pi$ pulse on the $\left|S,-\frac{1}{2}\right\rangle \rightarrow\left|D,-\frac{1}{2}\right\rangle$ transition using the 674 nm laser. These two pulses serve as the first $\frac{\pi}{2}$ pulse in a generalized Ramsey experiment, initializing the state $\left|\psi_{i}\right\rangle =\frac{1}{\sqrt{2}}\left(\left|D,-\frac{5}{2}\right\rangle +\left|D,-\frac{1}{2}\right\rangle \right)$. After preparing this initial state, a sequence of RF spin-echo pulses is repeated multiple times. Each spin echo pulse is preceded and followed by a wait time $\tau$, See Fig. 4b. The RF echo pulses are resonant with the Zeeman splitting of levels in the $D_\frac{5}{2}$, $f_{RF}=\Delta f_{Z}$, leading to precession of the spin $\frac{5}{2}$ in this manifold. The operation of this RF pulse is shown in Fig. 1c, and examples for this Rabi nutation under this large spin rotation are shown in Fig. 1d,1e. The RF pulse maps our ion from the $\left\{ \left|D,-\frac{5}{2}\right\rangle ,\left|D,-\frac{1}{2}\right\rangle \right\}$ sub-space to the $\left\{ \left|D,\frac{5}{2}\right\rangle ,\left|D,\frac{1}{2}\right\rangle \right\}$  sub-space and vice versa. Since the separation of the levels in each of these subspaces due to magnetic field is the same in magnitude and opposite in sign, the superposition phases accumulated in the wait times before and after each echo pulse due to a constant magnetic field cancel each other. Therefore, the superposition phase does not depend on the magnetic field to the first order. That is a direct consequence of the Zeeman splitting being linear with $m_{J}$. According to Eq. 1, the quadrupole shift depends on $m_{J}^{2}$, and therefore its contribution to the superposition phase has the same sign in the two subspaces, and hence added rather than canceled. The phase accumulated during each wait time $\tau$ is given by the difference between the quadrupole shift of the $m=\frac{5}{2}$ and $m=\frac{1}{2}$ states,
\begin{equation}
\Delta\phi_{arm}=\frac{9}{20\hbar}\frac{dE_{z}}{dz}\Theta\left(D,\frac{5}{2}\right)\left(3\mbox{cos}^{2}\left(\beta\right)-1\right)\tau.
\end{equation}
 Echo pulses with two wait times are repeated an even number of times, where the phase of the different RF pulses alternates between $0$ and $\pi$, to correct for pulse area imperfections. After the last $\tau$
  wait time, the ion is in a superposition $\left|\psi_{f}\right\rangle =\frac{1}{\sqrt{2}}\left(\left|D,-\frac{5}{2}\right\rangle +e^{i\phi_{total}}\left|D,-\frac{1}{2}\right\rangle \right)$
  where $\phi_{total}$ is the final superposition phase, and is equal to $\phi_{total}=2n\Delta\phi_{arm}$, where $n$ is the number of RF pulses. \par

 In order to detect this phase, we transfer the $\left|D,-\frac{5}{2}\right\rangle$ state to the $\left|S,-\frac{1}{2}\right\rangle$ state mapping $\left|\psi_{f}\right\rangle$ into $\left|\psi_{det}\right\rangle =\frac{1}{\sqrt{2}}\left(\left|S,-\frac{1}{2}\right\rangle +e^{i\phi_{total}}\left|D,-\frac{1}{2}\right\rangle \right)$. We measure $\phi_{total}$ by completing a Ramsey sequence and applying a $\frac{\pi}{2}$ pulse on the $\left|S,-\frac{1}{2}\right\rangle \rightarrow\left|D,-\frac{1}{2}\right\rangle$ transition with laser phase $\phi_{laser}$ with respect to the initial Ramsey $\frac{\pi}{2}$ pulse, followed by state-selective fluorescence at 422 nm. Note that the laser phase should be stable only between the two last detection pulses, and does not play any roll during the quadrupole phase accumulation time. The probability of measuring the ion in the D level, estimated using 300 repetitions per point, vs. $\phi_{laser}$, is shown in the inset of the top plot of Fig. 2. The accumulated phase $\phi_{total}$ is estimated by a maximum likelihood fit. In order to reject the phase accumulated during the pulses in the sequence (smaller than 0.1 rad), we subtract the phase measured in a reference experiment with $\tau=0$. \par

\begin{figure}
\centering
\includegraphics[width=9cm]{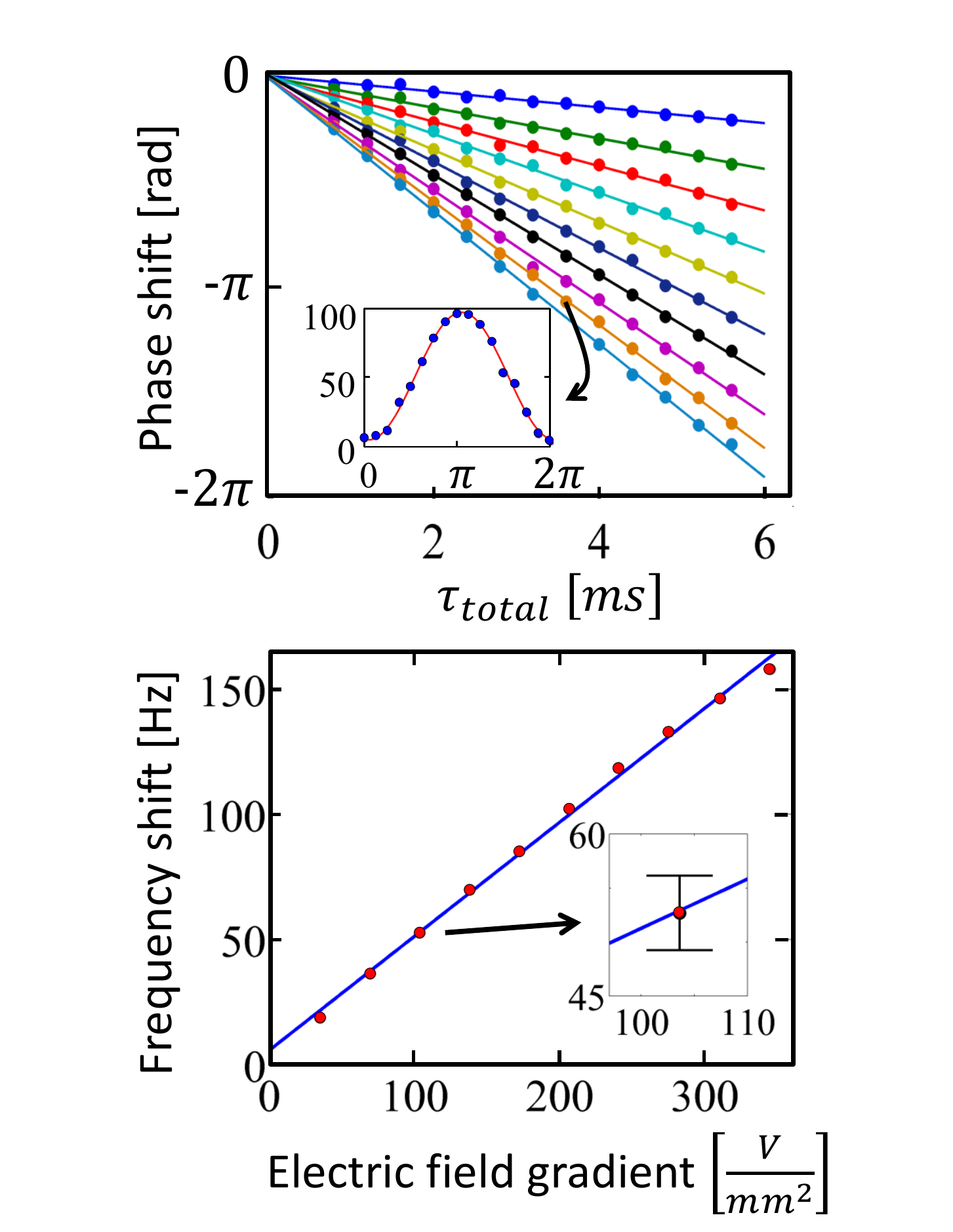}
\caption{Measured phase and frequency shifts as a function of the total experiment time $\tau_{total}$ and $\frac{dE_{z}}{dz}$\\ \textbf{Top:} Ramsey phase measurement as a function of the total experiment time, for different electric field gradients shown in different colors. Filled circles represent the phase of a Ramsey fringe fit. Solid lines show fitted linear curves to the data. As seen, the phase increases linearly with time. A typical Ramsey fringe is shown in the inset. \textbf{Bottom:} Quadrupole frequency shift as a function of the electric field gradient. Each point corresponds to the absolute value of a slope (in Hz) of each line in (a). 95\% confidence error-bars for one of the points are shown in the inset. The frequency shift is linear in the applied gradient within the measurement uncertainty. Solid line shows fitted linear slope to the data.}

\end{figure}

To measure the quadrupole moment, $\Theta\left(D,\frac{5}{2}\right)$, one has to determine both $\frac{dE_{z}}{dz}$ and $\beta$. The electric field gradient was estimated by measuring the ion harmonic frequency in the trap. Here, the ion motion was driven with an RF electric field while fluorescence due to excitation on the 422 nm transition was monitored. The RF frequency was scanned at a constant endcap voltage. Close to resonance the fluorescence rate is reduced due to Doppler shifts. The trap frequency was estimated from a fit of the fluorescence power vs. drive frequency. The gradient in the electric field is then given by $\frac{dE_{z}}{dz}=\frac{m\omega^{2}}{q}$ where $m$, $\omega$, $q$ are the mass of the ion, the trap frequency in the quadrupole axis and the charge of the ion respectively. The electric field gradient estimation was done for the same endcap voltages used on the quadrupole shift measurement, and had an error less than $0.1\%$. This error therefore contributes negligibly to the final quadrupole moment value compared with the statistical uncertainty in evaluating $\Theta\left(D,\frac{5}{2}\right)$ shown below. The contribution of the RF trap electric fields to the axial trap frequency was measured and subtracted from the electric gradient estimation. This contribution was less than $0.1\%$.

The top plot in Figure 2 shows the measured $ \phi_{total}$ as a function of $\tau_{total}$ for various electric field gradients. In all these measurements the magnetic field was aligned roughly at $45^{\circ}$ with respect to the quadrupole axis ($\beta \approx \pi/4$). Indeed the acquired phase varies linearly with $\tau_{total}$. The frequency shift was estimated by a maximum likelihood linear fit and is shown vs. electric field gradient in the top plot in Fig. 2. As seen, the frequency shift we measure is linear in the applied electric field gradient within our measurement error. \par

Instead of determining $\beta$ precisely, we measured the phase shift for different magnetic field directions. The magnetic field direction was varied by changing the relative current between the three Helmholtz coils. The main Helmholtz coil was held at a constant current. The current in a perpendicular coil was varied, and thus both the magnitude and the angle of the magnetic field were changed in a correlated manner. The magnetic field originating in these two coils was parallel to the optical table. We compensated magnetic field components perpendicular to the optical table using the third coil. Since the $z$ direction is parallel to the optical table to within $3^{\circ}$, the magnetic field was scanned in a plane containing the $z$ direction. We used the magnetic field sensitive transition $\left|S,-\frac{1}{2}\right\rangle \rightarrow\left|S,\frac{1}{2}\right\rangle$ to measure the magnetic field magnitude on the ion, and therefore also the angle of the magnetic field relative to the base angle defined by the main coil direction, $\beta_{0}$. The measured superposition phase as a function of $\beta$ is shown in Fig. 3. For each $\beta$ the superposition phase is seen to be linear with $\frac{dE_{z}}{dz}$. A linear fit reveals a small phase offset (always below $0.2 \mbox{rad}$ in magnitude) when $\frac{dE_{z}}{dz}=0$, which is removed from the entire data set for each $\beta$. Since both $\frac{dE_{z}}{dz}$ and $\tau$ are precisely known, we used a maximum likelihood fit to estimate both $\beta_{0}$ and $\Theta(D,5/2)$.

\begin{figure}
\includegraphics[width=9cm]{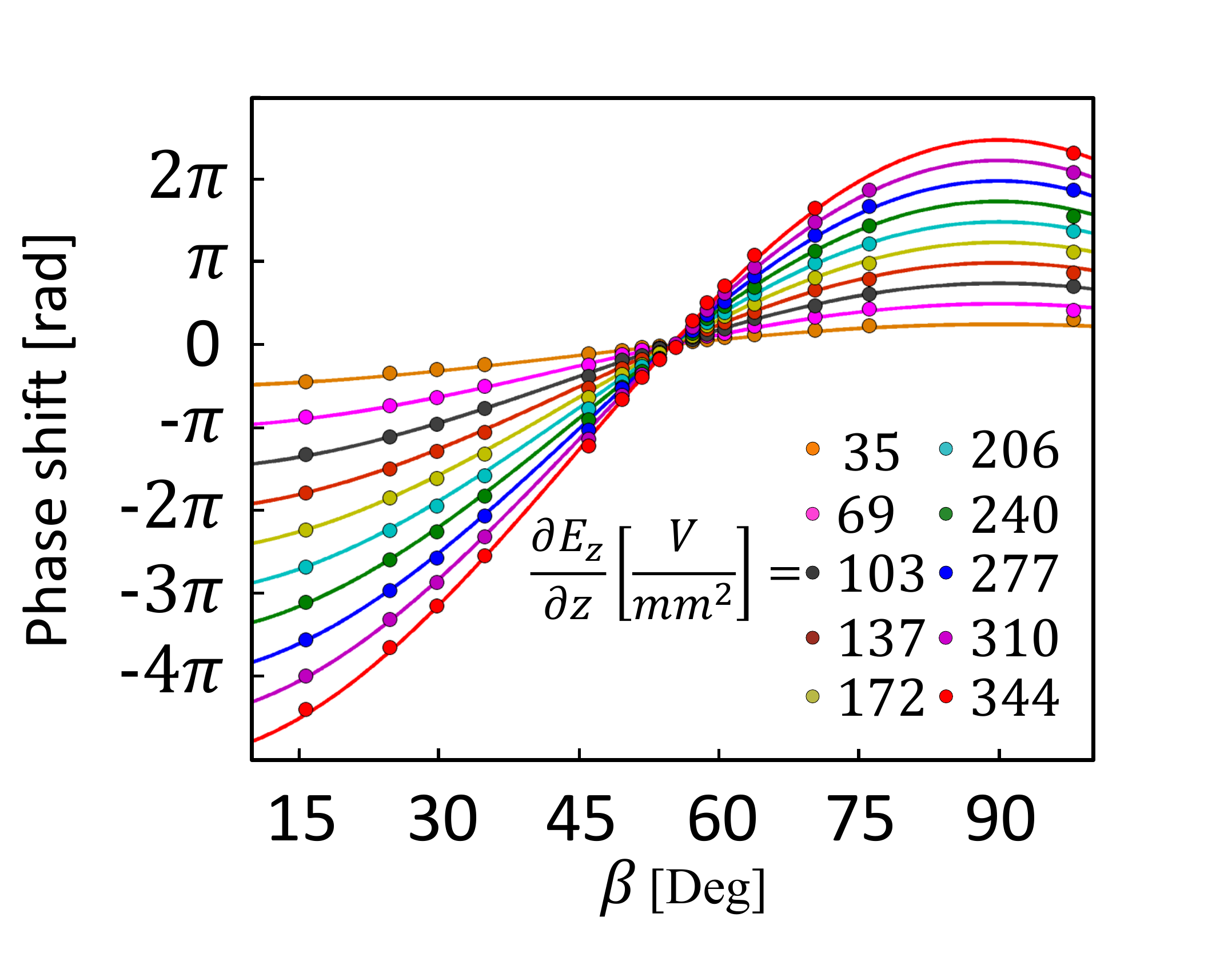}
\caption{The total accumulated phase due to the quadrupole shift, $\phi_{total}$, as a function of magnetic field angle with respect to the quadrupole axis - $\beta$, for different values of $\frac{dE_{z}}{dz}$. The total phase accumulation time for all points was $\tau_{total} = 4$ msec. Filled circles are measured phases. For each angle, the phase shift was linearly fitted and the phase offset for $\frac{dE_{z}}{dz}=0$ (always below 0.2 rad) was subtracted from the data set. . The error bars are too small to be seen on this scale. Solid lines are generated by a maximum likelihood fit to Eq. 3 taking the entire data set into account.}
\end{figure}

The measurement result for the quadrupole moment could be affected by several systematic effects. Since a second order Zeeman shift depends on $m_{J}^{2}$, it is not canceled by our sequence. The second order Zeeman shift in our experiment is $3.1\mbox{\ensuremath{\frac{MHz}{T^{2}}}}$, \cite{poli2014optical} and therefore in our case it would amount to less than $0.5 \mbox{\ensuremath{Hz}}$, in comparison to a few tens to hundreds of Hz of the quadrupole shift, and below our measurement accuracy. In fact, our sequence can serve as a tool to measure the second order Zeeman shift, by applying different magnitudes of magnetic field in a constant direction and measuring the superposition phase. Another type of systematic errors could arise from trap imperfections. Eq. 1 relies on the trap DC potential having perfect cylindrical symmetry, but the more general way of writing the potential that still satisfies Laplace equation is \\
$U=\frac{1}{4}\frac{dE_{z}}{dz}[(x^{2}+y^{2}-2z^{2})+\epsilon_{1}(x^{2}-y^{2})]$. Here we take into account a possible DC non-degeneracy in the radial frequencies ($\epsilon_{1}$). This term yields a phase proportional to $\sin^{2}(\beta)$ and it is small, since it reflects the deviation of the magnetic field direction from the plain defined by the trap axis and the direction $\vec{x}+\vec{y}$ where $\vec{x},\vec{y}$ are vectors parallel to the trap $x,y$ axes. We take $\epsilon_{1}$ term into account by adding it to the maximum likelihood fit for the final quadrupole moment value.

 The final result and its statistical $95\%$ confidence interval are $\Theta\left(D,\frac{5}{2}\right)=2.973^{+0.026}_{-0.033}\, ea_{0}^{2}$. \par

\begin{table}[h]
\includegraphics[width=8cm]{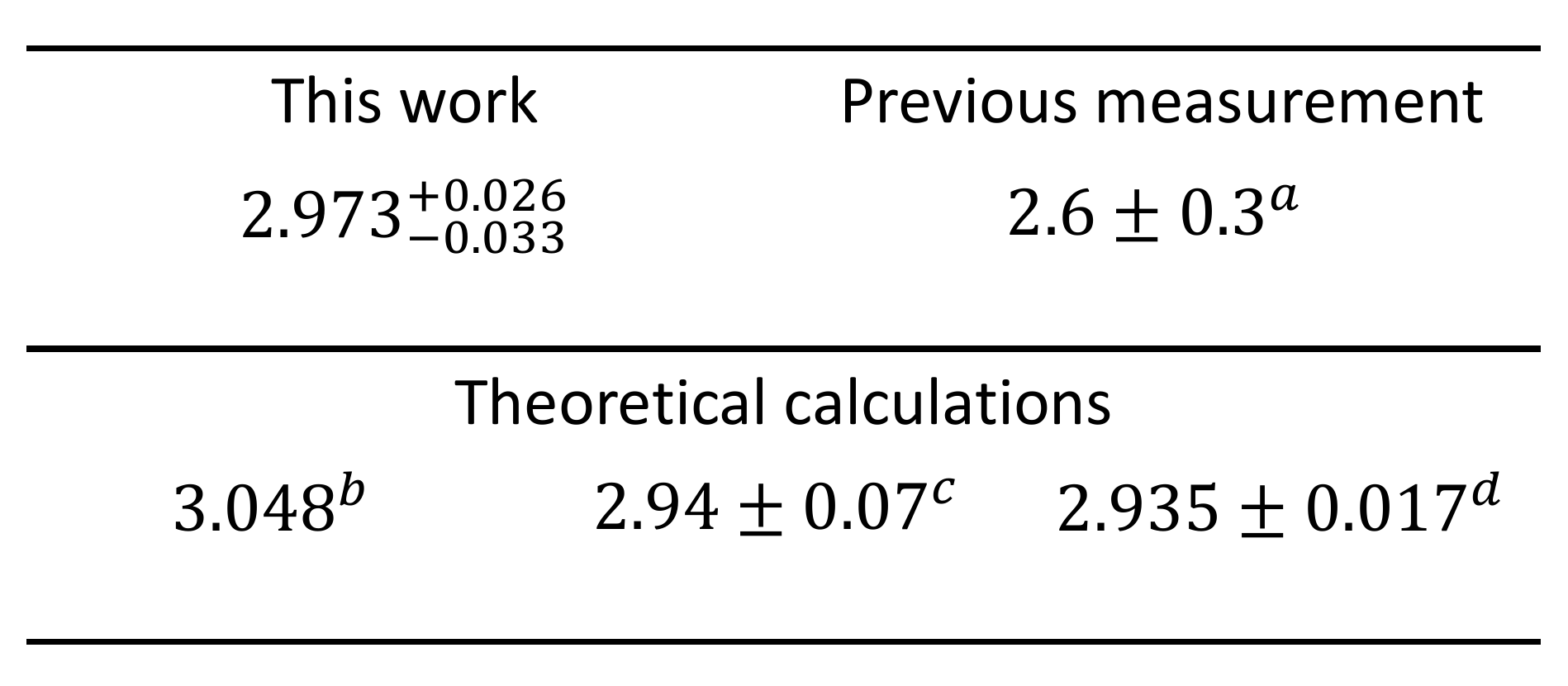}
\caption{A comparison between different measurements and calculations of the $4D_{5/2}$ level quadrupole moment in $^{88}Sr^{+}$. \textbf{(a)} \cite{barwood2004measurement}, \textbf{(b)} \cite{itano2006quadrupole}, \textbf{(c)} \cite{Mukherjee}, \textbf{(d)} \cite{Safronova}}
\end{table}

 The quadrupole moment of the $4D_{\frac{5}{2}}$ level in $^{88}Sr^{+}$ was previously measured by observing the shift of the optical clock transition at different electric field gradients \cite{barwood2004measurement}. Theoretical calculations of the quadrupole moment in $^{88}Sr^{+}$ were also carried out \cite{itano2006quadrupole,Mukherjee,Safronova}.
All these results are summarized in Table 1. Our measurement is in good agreement with the theoretical calculations, and is $1.2 \sigma$ away from the previous measurement with an uncertainty an order of magnitude smaller. \par

To conclude, we have developed a new method that combines dynamic decoupling together with the use of a multi-equidistant-level quantum probe to distinguish a small frequency shift that depends on $m_{j}^{2}$ from large noisy shifts that depend linearly on $m_{j}$. We thus measured the quadrupole shift of the $4D_{\frac{5}{2}}$ level in $^{88}Sr^{+}$ while decoupling it from magnetic field noise. Our method allowed us to measure the quadrupole moment of the $4D_{\frac{5}{2}}$ level with 1.1\% uncertainty. Our measurement verifies complicated many-body quantum calculations of atomic structure. In the future this method could be used to measure second order Zeeman shifts as well. since our method only requires a single ion and no entanglement, it can easily be used in order to place bounds on possible systematic shifts in precision optical spectroscopy. \par

We thank Tom Manovitz for many helpful discussions. This work was supported by the Crown Photonics Center, ICore-Israeli excellence center circle of light, the Israeli Science foundation, the Israeli Ministry of Science Technology and Space, and the European Research Council (consolidator grant 616919-Ionology)

\bibliographystyle{unsrt}	
\bibliography{quadrupole_shift_prl_template_version6}

\end{document}